# Quenching Dependence on Superconductivity in the Synthesizing Process of Single Crystals of $Rb_xFe_{2-y}Se_2$

Masashi TANAKA[1,2,*], Hiroyuki TAKEYA[2], and Yoshihiko TAKANO[2]

[1]*Graduate School of Engineering, Kyushu Institute of Technology, 1-1 Sensui-cho, Tobata, Kitakyushu 804-8550, Japan*

[2]*WPI-MANA, National Institute for Materials Science, 1-2-1 Sengen, Tsukuba, Ibaraki 305-0047, Japan*

*Corresponding author: Masashi Tanaka

Postal address: Graduate School of Engineering, Kyushu Institute of Technology

1-1 Sensui-cho, Tobata, Kitakyushu 804-8550, Japan

Tel/FAX: (+81)93-884-3204

E-mail: mtanaka@mns.kyutech.ac.jp

**Abstract**

Superconducting single crystals of Rb-intercalated FeSe compounds $Rb_xFe_{2-y}Se_2$ were prepared by using a starting material of $Rb_2Se$ as a Rb source. The superconducting properties and the surface microstructures were systematically controlled by varying the cooling rate in the quenching process. The higher cooling rate in the sample provided a higher superconducting transition temperature with highly connected superconducting mesh-like surface structure. Extremely slow-cooling process led to the complete isolation between the superconducting domains.

Keywords

A. Iron-based Superconductors; A. Rb intercalated FeSe; B. One-step preparation with quenching

# 1. Introduction

After the discovery of superconductivity in potassium intercalated FeSe [1], tremendous progress has been made in the study of related $A_x$Fe$_{2-y}$Se$_2$ ($A$ = K, Rb, Cs, Tl/Rb, Tl/K) systems which exhibit relatively high superconducting (SC) transition temperatures $T_c$ > 30 K [2-7]. By an appropriate tuning of the basic FeSe layer structure, SC phase with $T_c$ ~ 40-48 K was also observed in several research groups [8-13].

However, many studies have shown that intrinsic phase separations occur in the material, leading to the coexistence of a SC phase with the ThCr$_2$Si$_2$-type structure (122-phase) and an Fe-vacancy ordered insulating $A_2$Fe$_4$Se$_5$ phase with $\sqrt{5} \times \sqrt{5} \times 1$ superstructure (245-phase) [14]. This multiphase nature of these compounds has been made it difficult to understand the mechanisms responsible for high-temperature superconductivity. These complications about superconductivity in K-Fe-Se system has begun to solve in the recently days by means of several kinds of unique techniques [15-21]. The 245-phase in K$_x$Fe$_{2-y}$Se$_2$ may become a driving force for the growth of the higher-$T_c$ phase. Anyhow, a development of the appropriate synthetic method to obtain the pure superconducting phase in these intercalated FeSe-based compounds is still urgently required.

In the case of rubidium structures, there has been reported that it also shows complex mixture of phases due to intrinsic phase separation at specific transition temperatures around 200-250°C [22,23]. And the superconducting properties and microstructure are known to show strong dependence upon post-annealing [24-27]. However, it is difficult to control the Rb stoichiometry because of high reactivity of Rb metal, still more in the application of post-annealing. It is useful that if we know how the superconducting properties can be controlled by modifying the synthesis conditions, and if we establish a method for controlling the superconducting properties systematically in a one-step way without post-annealing.

In this study, we have found such a route for obtaining superconducting $Rb_xFe_{2-y}Se_2$ single crystals by using $Rb_2Se$ as a Rb source. And the crystals prepared by various quenching conditions were evaluated its SC properties and surface morphology. This is a part of studies to establish Big Data resources in a variation of synthesizing process for data mining in the development of material science.

## 2. Experimental

$Rb_xFe_{2-y}Se_2$ single crystals were prepared using similar way to a one-step method with quenching [20,28]. Starting powder of $Rb_2Se$ was firstly prepared by reacting Rb metal and Se at 150-200°C [29]. Powders of $Rb_2Se$, Fe, and Se grains were mixed with a nominal composition of $Rb_{0.8}Fe_2Se_2$ in an Ar atmosphere. The mixtures were placed in an alumina crucible, and were sealed in evacuated quartz tubes. The quartz tube was heated to 1030°C in 11 h, the temperature was then held for 3 h, followed by slowly cooling to room temperature at a rate of 6°C/h (slow-cool, Sample A), or cooling to a quenching temperature of 700°C (quench, Sample B-D). The quenching was done by dropping the hot quartz tube into 20°C water stored in a small container (Sample B), flowing tap-water (Sample C), and a bucketful of iced-water (Sample D). The cooling rate normally depends on its temperature difference between the sample and refrigerant, then the cooling rate in Sample D is expected to be the highest, followed by Sample C, B, A. The single-crystal X-ray diffraction experiments at room temperature were carried out by using a three-circle diffractometer with a CCD area detector (Smart Apex II, Bruker). Each crystal was sealed into an evacuated thin quartz capillary during the diffraction measurement. The crystal structure was solved and refined by using the program

SHELXT and SHELXL [30,31], respectively, in the WinGX software package [32]. The temperature dependence of magnetization was measured using a SQUID magnetometer (MPMS, Quantum Design) down to 2 K under a field of 10 Oe, and the field was applied parallel to the *ab*-plane. The temperature dependence of electrical resistivity was measured down to 2.0 K, with a Physical Property Measurement System (PPMS, Quantum Design) using a standard four-probe method with constant current mode. The electrodes were attached in the *ab*-plane with silver paste. Back scattered electron (BSE) images and energy-dispersive X-ray (EDX) spectra were observed using a scanning electron microscope (JEOL, JSM-6010LA).

## 3. Results and discussions

*3.1 Single Crystal Growth and Structural Characterization*

In the course of the synthetic research, a series of samples in the Rb-Fe-Se system have been obtained by direct reaction of Rb metal and FeSe powder with varying the nominal composition. Literatures reported that the superconducting phase is required to adjust the compositional ratios to a narrow range [22,23]. However, it is difficult to control and reproduce the products because of high reactivity of Rb metal, reflecting the subtle interplay of kinetic and thermodynamic factors, such as the cooling rate, the internal pressure inside the ampule, and the initial ratio of the reagents. Then we firstly prepared $Rb_2Se$ as a starting material for the single crystal. It would lead to easy handling of nominal composition and systematic synthesis, then also would be possible to control the superconducting properties.

The single crystals prepared by several cooling processes described in the experimental section were characterized by the single-crystal X-ray diffraction experiment at room temperature. The typical single crystal refinement for the Sample B converged to the $R1$ values of 6.57% for $I \geq 2\sigma(I)$ with the 245-phase structure. The anisotropic displacement parameters were all positive and within similar ranges, all of which appeared to be physically reasonable. The crystallographic details are listed in the supplementary material and attached as a Crystallographic Information File (CIF).

*3.2 Surface Morphology*

The microstructure of the surface of the single crystals shows unique morphology depending on its cooling condition as shown in Fig. 1. It can be clearly seen that all samples include two-phases with alternating dark and bright regions. The composition of the white contrasted regions in Fig. 1(a) shows stoichiometric values related to $Rb_{0.72(2)}Fe_{1.63(3)}Se_2$. The Fe to Se ratio is close to 4:5 and it is similar values to the "out-of-island" phase in the microsamples of [21], suggesting that this phase corresponds to an insulating 245-phase, $Rb_2Fe_4Se_5$. On the other hand, the dark contrasted island-like regions have a composition of $Rb_{0.51(2)}Fe_{1.90(3)}Se_2$. The island-like regions in the slow-cooled crystal contain more iron and less rubidium compare to that found in the white contrasted region in the same crystals. This tendency in compositional ratio is similar to that in the case of K-Fe-Se system, except for the opposite contrast reflecting the difference in the atomic numbers. In the quenched crystals of Figs. 1(b)-(d), the dark contrasted regions become a mesh-like network, namely Fe-rich parts are linked in entire region of the surface of the quenched crystals. The mesh-like network becomes finer and denser with increase of the cooling rate in the sequence of Sample B to D (from Fig. 1(b) to Fig. 1(d)). The area ratio of the dark domains in Fig. 1 were estimated to be (a) ~15%, (b) ~25%, (c) ~28%, and (d) ~30%, respectively, by using calculation based on the binarized BSE images with the same way to [33]. A faster cooling rate leads to a suppression of the formation of the 245-phase in exchange for the high connectivity in the dark domains.

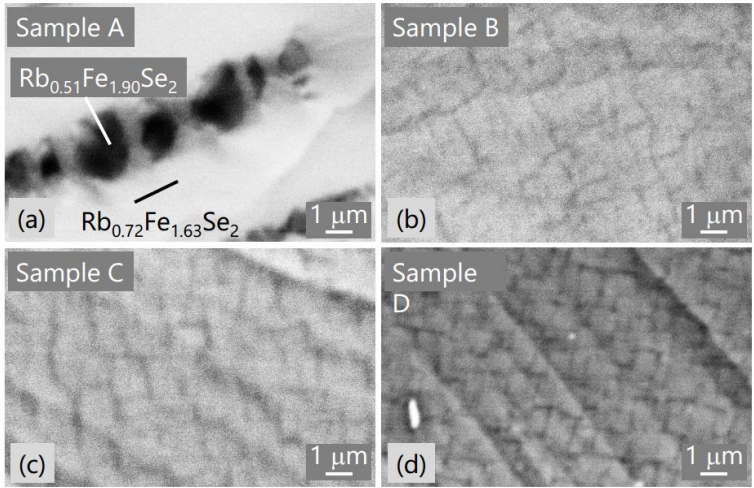

Figure 1. BSE images of the single crystals obtained by various quenching conditions.

*3.3 Superconducting Properties*

The superconductivity in obtained single crystals considerably differs related to the connectivity and density of the dark contrasted regions in Fig. 1, namely it was strongly affected by the cooling rate in the synthesizing process. Figure 2 shows the temperature dependences of the magnetization for the four different quenching conditions. Sample D shows a superconducting transition around $T_c$ ~30 K with a large shielding volume fraction. The $T_c$ and shielding fraction tends to decrease with decrease of the cooling rate.

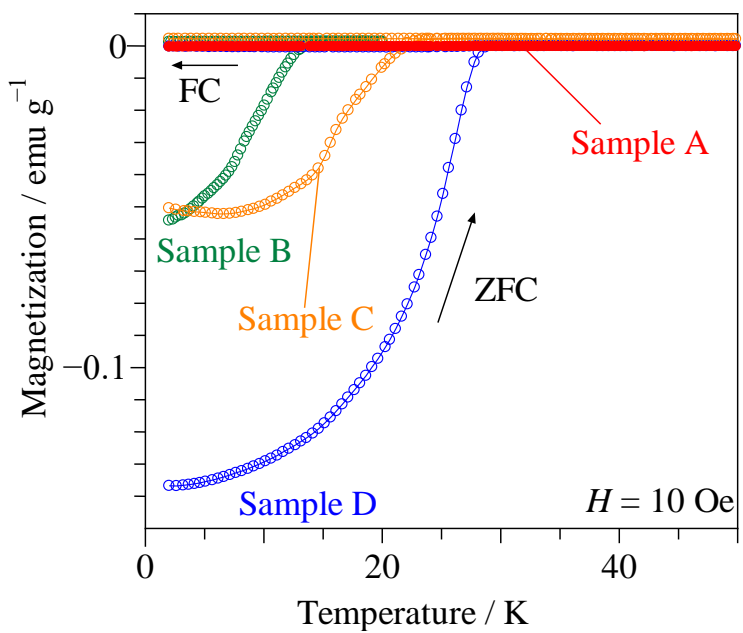

Figure 2. Temperature dependence of the magnetization of single crystals quenched in various conditions.

As the cooling rate increases, the electrical resistivity at 300 K decreases along Samples A > B > C > D with suppressing the semiconducting behavior in the normal state conduction region (Fig. 3). Sample A shows a broad hump at around 200 K and the broad hump like structure gradually moves to the higher temperature side in the Samples, B, C, D.

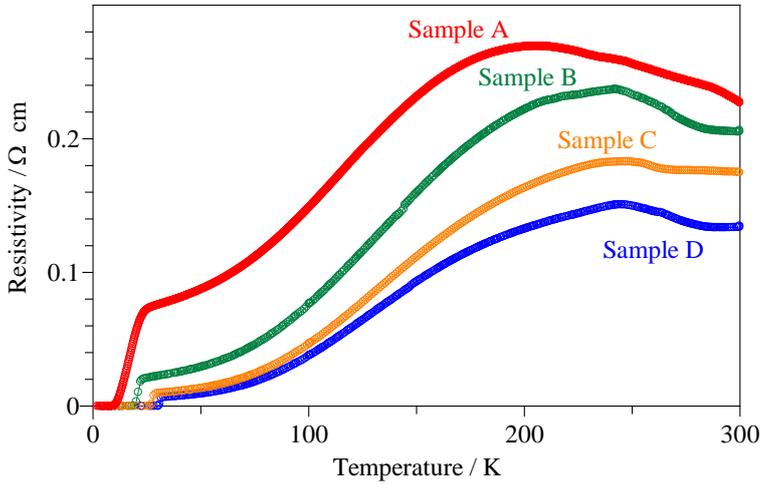

Figure 3. Temperature dependences of resistivity of single crystals obtained by various quenching conditions.

All the samples show resistivity drop attributed to the superconducting transition in the temperature region below ~40 K as shown in Fig. 4. In the enlargement scale of the transition temperature, Sample D shows a clear resistivity drop at ~32 K and zero-resistivity at ~30 K, which clearly corresponds to the Meissner signal in the magnetization measurement. The transition temperatures become lower along Samples D > C > B > A, which is evidently related to the decrease of the cooling rate in the quenching process. Apparently strange zero-resistivity in Sample A may be attributed to the three dimensionally connected superconducting paths as is observed in the elemental distribution in a cross-sectional plane of microsamples [21].

The resistivity was measured under various magnetic fields along the *ab*-plane (*H*//*ab*) and the *c*-axis (*H*//*c*) as shown in Figs. 4(a) and (b), respectively. The superconductivity in each sample is suppressed by applying a magnetic field both in the directions along *H*//*ab* and *H*//*c*. It was largely suppressed with increasing field parallel to *c*-axis in comparison to that in the case of *ab*-plane, which may be attributed to the layer structured nature in the samples.

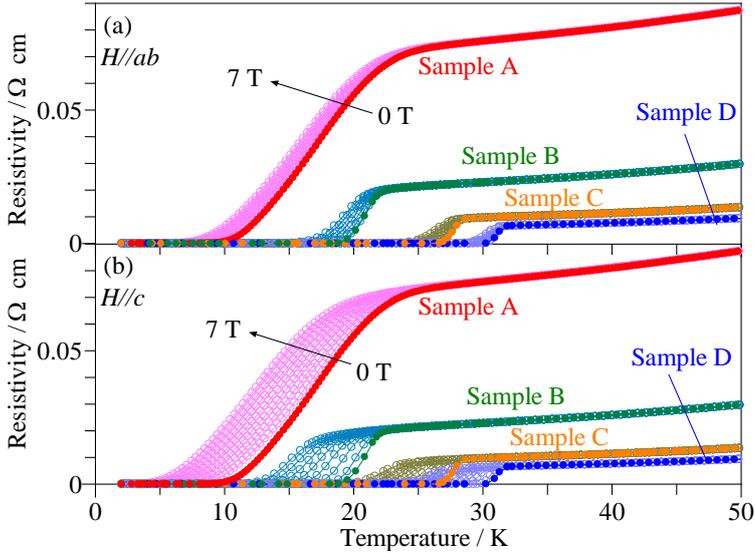

Figure 4. Enlarged scale of resistivity under various magnetic fields parallel to *ab*-plane (a) and *c*-axis (b).

Figure 5 shows the temperature dependence of the upper critical field ($H_{c2}(T)$), estimated using the criterion of the 90% drop of the normal state resistivity. The $H_{c2}(0)$ in *H*//*ab* and *H*//*c* in the Sample D were estimated to be $H_{c2}^{//ab}(0)$ = 151.6 T and $H_{c2}^{//c}(0)$ = 42.4 T, respectively, from the Werthamer-Helfand-Hohenberg (WHH) approximation for the Type II superconductor in a dirty limit [34]. The superconducting parameters of the other obtained samples are summarized in Table I. From these results, the superconducting anisotropic parameter $\gamma = H_{c2}^{//ab}(0) / H_{c2}^{//c}(0)$ is determined to be 3.4-3.7. The $T_c$, $H_{c2}(0)$ and $\gamma$ values obtained in Sample D are comparable to those of the post-annealed $Rb_xFe_{2-y}Se_2$ crystals [25], even though the crystals are prepared without annealing. Above all results suggest that the synthesizing route using $Rb_2Se$ is quite effectively control the superconductivity in $Rb_xFe_{2-y}Se_2$ single crystals.

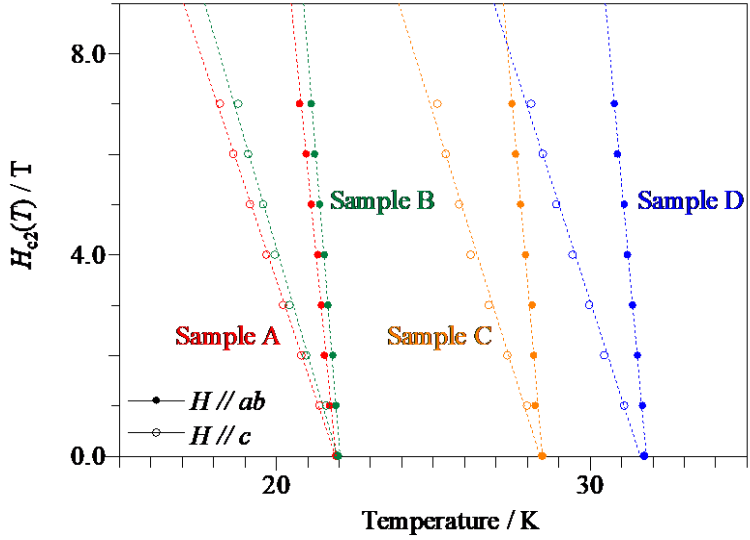

Figure 5. Temperature dependence of upper critical field $H_{c2}(T)$ obtained crystals in different applied magnetic field directions. The dotted lines show WHH approximation.

Table I. Superconducting parameters in the single crystals of $Rb_xFe_{2-y}Se_2$.

|  | Sample A | Sample B | Sample C | Sample D |
|---|---|---|---|---|
| Coolant | (Slow-cooled) | (Stored water) | (Flowing tap-water) | (Iced water) |
| $T_c^{onset}$ (K) | 21.9 | 22.0 | 28.5 | 31.7 |
| $H_{c2}^{//ab}(0)$ (T) | 95.7 | 118.2 | 143.2 | 151.6 |
| $H_{c2}^{//c}(0)$ (T) | 28.2 | 32.4 | 39.4 | 42.4 |
| $\gamma$ | 3.4 | 3.7 | 3.6 | 3.6 |

*3.4 In the case of "much" slower cooling rate*

When the starting mixture was kept at 200°C during the slow-cooling process, such a "super" slow-cooling brought a different character with the former slow-cooled crystals (Sample A). Figure 6(a) shows a surface microstructure of "super" slow-cooled crystals. It is apparently the same morphology with that of Sample A. However, the magnetization and resistivity measurements showed quite different characters from Sample A as shown in Figs. 6(b), (c). The resistivity showed semiconducting behavior, even though the magnetization showed a clear superconducting transition around ~32 K. Since 200°C is a temperature that Fe atoms start to diffuse in the crystal structure, keeping long time at 200°C causes a sufficient Fe exclusion from the 245-phase [20]. It would lead to considerable concentration of Fe onto the island like parts, resulting in the relatively higher $T_c$ compare to Sample A [20]. The results in magnetization and resistivity imply that the superconducting regions are completely embedded within an insulating continuum. The Arrhenius plot represents a liner correlation between the resistivity and the inverse of temperature above 100 K, indicating the behavior is thermal activation type (inset of Fig. 6(c)). The activation energy was estimated to be ~0.08 eV. Below 100 K, the resistivity starts to show the variable range hopping type behavior. These conduction behaviors seem to be in "granular insulator regime" in a picture of nano-sized granule superconductivity in $Rb_xFe_2Se_2$ with randomly dispersed within the insulating $Rb_2Fe_4Se_5$ matrix [35]. These electronic conducting phenomena including Samples A-D might be well analyzed and described by the manifestation of granularity. The granularity discussion will be appeared somewhere [36].

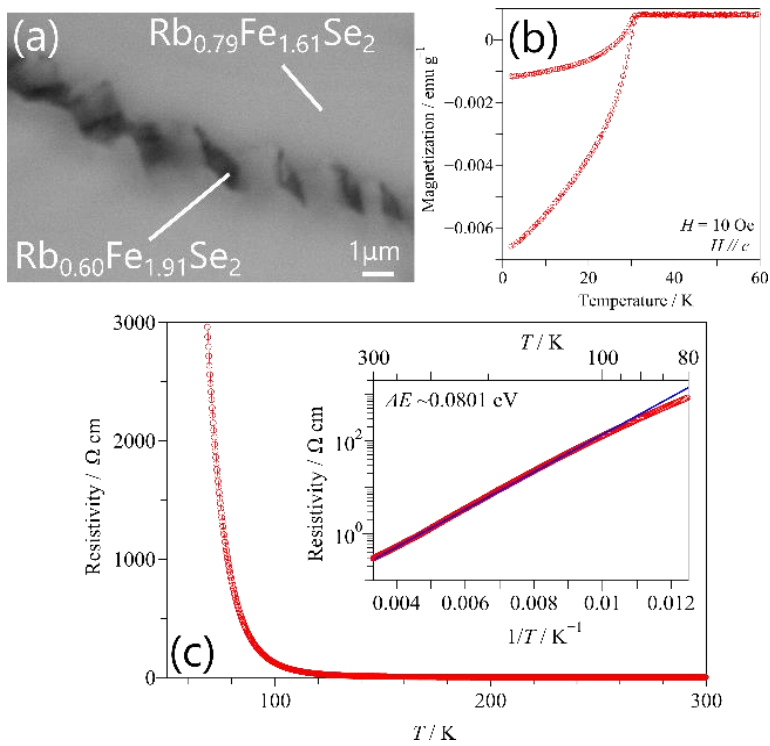

Figure 6. BSE image (a), temperature dependence of magnetization (b) and resistivity (c), of single crystals $Rb_xFe_{2-y}Se_2$ obtained extremely slow-cooling. The inset in (c) is the Arrhenius plot of the resistivity as a function of the reciprocal temperature.

## 4. Conclusion

Superconducting single crystals of Rb-intercalated FeSe compounds $Rb_xFe_{2-y}Se_2$ were easily prepared by using a starting material of $Rb_2Se$ as a Rb source. The superconducting properties and surface microstructures can be systematically controlled by varying the conditions of quenching process. The faster quenching rate produced a higher superconducting transition temperature with highly connected superconducting mesh-like region in the surface morphology. Extremely slow-cooling process led to the complete isolation between the superconducting region. The resistivity behavior including the normal state may be well described in the manifestation of granularity.

**Acknowledgments**

This work was partially supported by JST CREST Grant No. JPMJCR16Q6.